\documentclass[10pt, aps, prl, superscriptaddress, nofootinbib,
               amsmath, 
               twocolumn, 
               showpacs, 
               floatfix, 
               ]{revtex4-2}

\providecommand{\exclude}[1]{}

\usepackage{etoolbox}
\newtoggle{mypaper}
\toggletrue{mypaper}
\iftoggle{mypaper}{
  \usepackage[aps, arXiv, todo, secnoindent]{my_paper}
  \iftoggle{arXiv}{
    \newcommand{\EPAPS}{\nameref{sec:supp}}
  }{
    \newcommand{\EPAPS}{\cite{EPAPS}}
  }
}{
  \usepackage{graphicx}
  \usepackage{booktabs}
  \graphicspath{{./figures/aps/}{./figures/}}

  \usepackage{siunitx}
  \sisetup{
    round-mode = none,   
    mode = text,         
    detect-all = true,            
    detect-mode = false,          
    list-units = single,  
    unit-mode = text,
    number-mode = math,
  }

  \usepackage[nameinlink]{cleveref}
  
  \providecommand{\mycaptionhook}{}
  \let\oldcaption\caption
  \renewcommand{\caption}[2][]{\oldcaption[#1]{\mycaptionhook{}#2}}

  \usepackage[version=3]{mhchem}

  \providecommand{\mysc}[1]{\MakeUppercase{#1}}
  \providecommand{\upsc}[1]{\MakeUppercase{#1}}
  
  \newcommand{\EPAPS}{\cite{EPAPS}}
}

\usepackage{my_acronyms}

\usepackage[normalem]{ulem}

\usepackage[caption=false]{subfig}
\newcommand{\labelphantom}[1]{\subfloat{\label{#1}}}

\graphicspath{{figures/}}

\newcommand{\abs}[1]{\ensuremath{\lvert{#1}\rvert}}
\newcommand{\ket}[1]{\ensuremath{\lvert{#1}\rangle}}

\newcommand{\pdiff}[3][]{\frac{\partial^{#1} #2}{\partial{#3}^{#1}}}
\newcommand{\vect}[1]{\vec{#1}}
\newcommand{\I}{\mathrm{i}}
\renewcommand{\d}{\mathrm{d}}
\newcommand{\mat}[1]{\boldsymbol{#1}}
\newcommand{\op}[1]{\hat{#1}}
\newcommand{\ua}{2,0}
\newcommand{\ub}{1,0}

\DeclareMathOperator{\Ai}{Ai}
\DeclareMathOperator{\Bi}{Bi}
\DeclareMathOperator{\Ci}{Ci}

\newcommand{\cl}{\mathrm{cl}}

\DeclareSymbolFont{greekletters}{OML}{zplm}{m}{it}

\usepackage[inline]{enumitem}
\setlist[enumerate]{label=\alph*)}%
\appto\mycaptionhook{%
  \setlist[enumerate]{label=\textbf{\alph*})}%
  \crefname{enumi}{panel}{panels}%
  \Crefname{enumi}{Panel}{Panels}%
}

\begin{document}

\title{Atom Interferometric Imaging of Differential Potentials Using an Atom Laser}

\author{M.~E.~Mossman}
\affiliation{Department of Physics and Biophysics, University of San Diego, San Diego, California 92110, USA}
\affiliation{Department of Physics and Astronomy, Washington State University, Pullman Washington 99164, USA}

\author{Ryan~A.~Corbin}
\affiliation{Department of Physics and Astronomy, Washington State University, Pullman Washington 99164, USA}

\author{Michael McNeil Forbes}
\email{m.forbes@wsu.edu}
\affiliation{Department of Physics and Astronomy, Washington State University, Pullman Washington 99164, USA}
\altaffiliation{Department of Physics, University of Washington, Seattle Washington 98105 USA}

\author{P.~Engels}
\email{engels@wsu.edu}
\affiliation{Department of Physics and Astronomy, Washington State University, Pullman Washington 99164 USA}

\begin{abstract}
  \noindent 
   Interferometry is a prime technique for modern precision measurements. 
   Atoms, unlike light, have significant interactions with electric, magnetic, and gravitational fields, making their use in interferometric applications particularly versatile. 
   Here, we demonstrate atom interferometry to image optical and magnetic potential landscapes over an area exceeding $\SI{240}{\micro m}\times\SI{600}{\micro m}$. 
   The differential potentials employed in our experiments generate phase imprints in an atom laser that are made visible through a Ramsey pulse sequence. 
   We further demonstrate how advanced pulse sequences can enhance desired imaging features, e.g.\ to image steep potential gradients.
   A theoretical discussion is presented that provides a semiclassical analysis and matching numerics. 

\end{abstract}

\maketitle
\clearpage

\paragraph{Introduction}
From optical precision measurements~\cite{Yang_2018} to detecting minute ripples in space-time~\cite{Abbott:2016}, interferometry is a keystone of modern science.
Compared to their light-based counterparts, matter-wave interferometers have some decisive advantages~\cite{Bongs_2019}, including significant sensitivity to electric, magnetic, and gravitational fields~\cite{Robins_2013,Rosi_2014,Overstreet:2022}.
Access to both motional and internal degrees of freedom makes atom interferometers suitable for many applications, including timekeeping with ultracold atoms~\cite{Cronin2009}, inertial measurements~\cite{Abend:2019,Stray_2022}, and fundamental studies of quantum dynamics~\cite{Moulder_2012,Corman_2014,Eckel_2014,Del-Pace:2022}.

Here, we demonstrate the two-dimensional imaging of differential potentials based on atom interferometry with an atom laser -- a coherent stream of atoms.
Atom lasers~\cite{Mewes1997, Naraschewski1997, Ketterle:1997, Steck1998,Bloch1999, Schneider1999, Ballagh:2000, Bloch:2000, LeCoq2001, Bloch:2001, Chikkatur2002, Haine:2002, Lee:2015, Harvie:2020, Riou:2008, Mossman:2021} can be generated by coherently outcoupling atoms from a trapped dilute-gas \gls{BEC} into an untrapped quantum state, creating a two-dimensional sheet of atoms in an accelerated reference frame.
Our interferometric imaging technique employs a Ramsey pulse sequence~\cite{Ramsey_1950,Bransden:2003}: two sequential $\pi/2$ pulses, via coherent microwaves, separated by a wait time.
The resulting images reveal contourlike lines of an applied differential potential.
The potential causes a phase imprint that can be measured across the entire atom laser in a single run of the experiment,  extending over an area exceeding $\SI{240}{\micro m} \times \SI{600}{\micro m}$.
Unlike previous atom interferometric work with a pulsed-output atom laser~\cite{Doering_2009} or with thermal atoms ~\cite{Ramola:2021}, our work utilizes a quasicontinuous atom laser to map out two-dimensional potential landscapes.
We demonstrate this technique with two types of differential potentials: a magnetic field that acts differently on two hyperfine states due to the Zeeman effect, and an optical dipole potential that is tuned to be attractive for one hyperfine state and repulsive for another.

\begin{figure}[htb!]
  \centering
  \labelphantom{fig:setupa}
  \labelphantom{fig:setupb}
  \labelphantom{fig:setupc}
  \labelphantom{fig:setupd}
  \labelphantom{fig:setupe}
  \includegraphics[width=\linewidth]{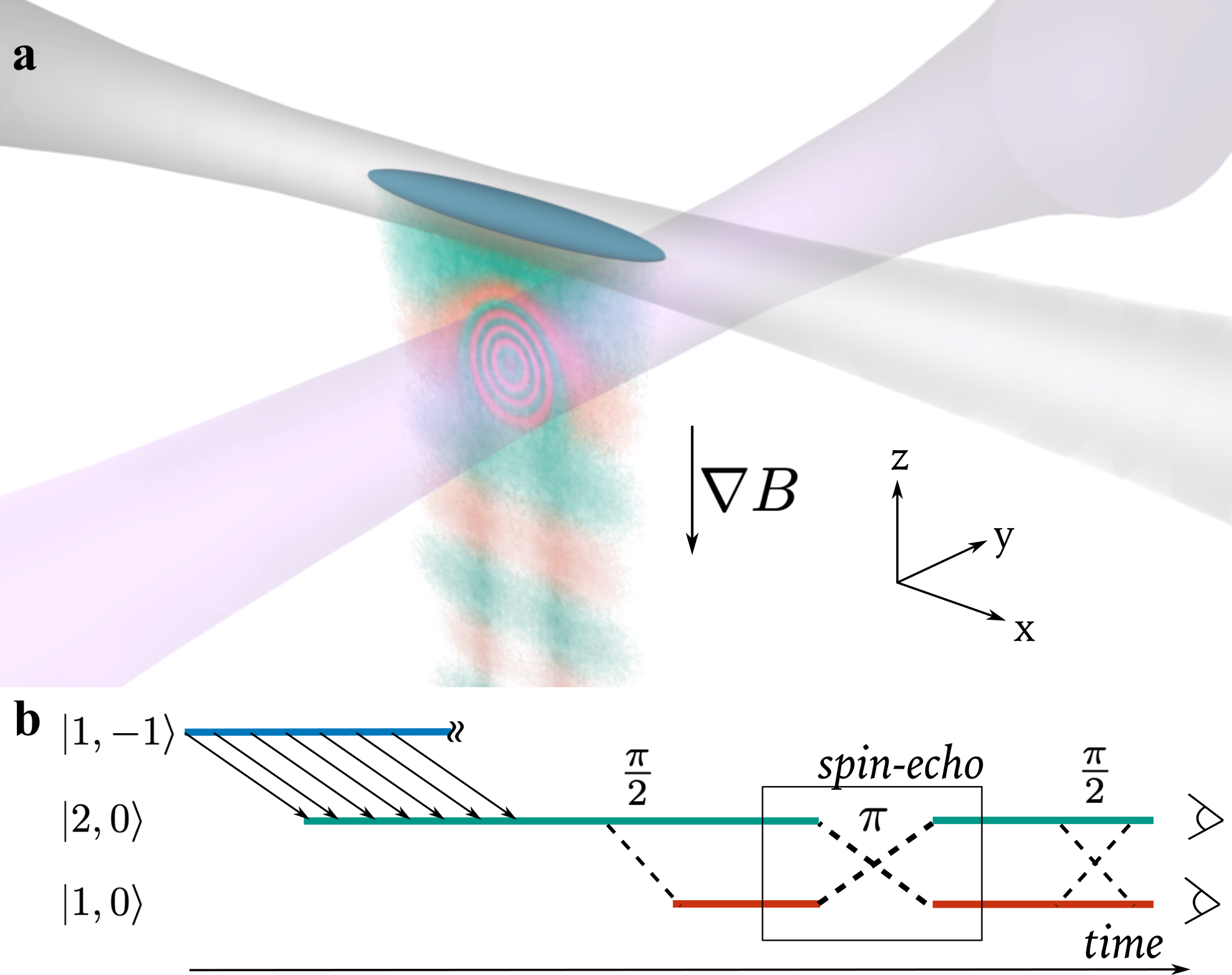}\\[1ex]
  \includegraphics[width=\linewidth]{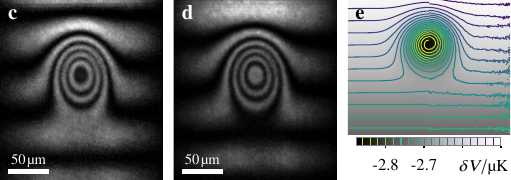}
  \vspace{-1.75\baselineskip}   
  \caption{\begin{enumerate*}
    \item\label{item:a}%
      Experimental setup: a \gls{BEC} (blue) of \ce{^{87}Rb} atoms in the $\ket{F, m_F} =\ket{1, -1}$ state is held in a dipole trap (gray).
      Atoms transferred to the $\ket{2, 0}$ state are accelerated out of the trap in the presence of gravity and a magnetic gradient, generating an atom laser that travels downward.
      An additional optical laser (violet) crosses the atom laser to generate weak differential potentials.
    \item\label{item:b}%
      Interferometric imaging procedure.
      A first Ramsey pulse places the whole extent of the atom laser into a superposition state.
      Optionally, a $\pi$ pulse can be inserted for realizing a spin echo.
      After a wait time, a second Ramsey pulse closes the interferometer with an arbitrary controllable phase, e.g.,\ %
    \item $\theta=0$ and
    \item $\theta=\pi$.
    The images in \textbf{c)} and \textbf{d)} are taken with a Ramsey pulse spacing of \SI{0.70}{ms} and have each been averaged over ten independent runs of the experiment.
    \item By directly fitting the interference patterns (left), or with phase retrieval techniques (right), we can accurately reconstruct the differential potential $\delta V(x, z)$.
    See \EPAPS{} for details.
    \end{enumerate*}
  }
  \label{fig:setup}
  \vspace{-.25\baselineskip} 
\end{figure}

As a practical application, we image a magnetic quadrupole field present in our experimental chamber, and show how a variation of the imaging pulse sequence can enhance desired features. 
Experimental results are well described by a semiclassical theory. 
Future applications include material science studies in hybrid quantum systems~\cite{Taylor:2021}, studies of interaction effects in quantum caustics~\cite{ODell:2012, Mumford:2017, Mumford:2019}, and branched flow~\cite{Heller:2021}.

\paragraph{Experimental procedure and results}
Our experiments begin with a dilute-gas \gls{BEC} of $\sim \num{4e6}$ atoms of \ce{^{87}Rb} in the $\ket{F,m_F} = \ket{1,-1}$ hyperfine state.
The \gls{BEC} is held in a hybrid trap formed by a focused infrared laser that provides mostly radial confinement, and a quadrupole magnetic field that provides additional support against gravity and axial confinement [\cref{fig:setupa}].
The resulting harmonic trap frequencies are $\{\omega_x,\, \omega_y,\, \omega_z\} = 2\pi\times\{3.7,\, 39.7,\,30.1\}~\si{Hz}$, with the weakly confined $x$ axis directed horizontally in the images.
From this trapped \gls{BEC}, a coherent stream of atoms is outcoupled to form an atom laser by using microwave radiation that gradually transfers atoms to the $\ket{2,0}$ state, which is only weakly supported by the magnetic gradient. These transferred atoms fall out of the trap, accelerating downward from the injection site.
After \SI{10}{ms} of continuous outcoupling from the \gls{BEC}, a brief \SI{68}{\micro s}-long microwave pulse puts the entire atom laser into a coherent superposition of the $\ket{2,0}$ and $\ket{1,0}$ state.

The falling atoms can be further manipulated with a dipole potential created by impinging focused laser light that is detuned from a resonant transition.
Under appropriate conditions, this can lead to strong mechanical effects such as intricate patterns of caustics~\cite{Mossman:2021}.
The sign and strength of the dipole potential depend on the intensity and wavelength of the laser in relation to the resonance lines of the atom.
Here, we exploit this versatility by choosing a laser wavelength such that the resulting potential is attractive for atoms in the $\ket{1,0}$ state and repulsive for atoms in the $\ket{2,0}$ state. 
See~\EPAPS{} for more details.
The dipole potential focuses or defocuses the $\ket{1,0}$ or $\ket{2,0}$ atom lasers respectively, and can in both cases form caustics for sufficiently high powers~\cite{Mossman:2021}.
Here we use weak dipole strengths that have only small mechanical effects on the atom laser, probing mostly the so-called ``Aharonov-Bohm'' phase~\cite{Overstreet:2022}.
The dipole is centered $z_{d} = \SI{89.9(4)}{\micro m}$ below the injection site, and the Gaussian waist radius of the dipole laser $w=\SI{38.9(6)}{\micro m}$ is smaller than the transverse extent of the atom laser ($\approx\SI{240}{\micro m}$).

In addition to the dipole potential, a differential potential for atoms in the $\ket{1,0}$ and $\ket{2,0}$ states can be generated by a magnetic field, shifting their energies by the quadratic Zeeman effect.
Here, with a background field of approximately \SI{10}{G} and a vertical gradient of $\d{B}/\d{z}=\SI{-25.10\pm 0.01}{G/cm}$, the energy of the $\ket{2,0}$ ($\ket{1,0}$) state increases (decreases) with $h \times \SI{12.43\pm0.06}{kHz/mm}$ in the $-z$ direction, determined using a linear approximation across the region of magnetic fields covered by the atom laser (see~\EPAPS{} for details).
This magnetic field is present in the full region of the atom laser, whereas the dipole laser is focused to a region just below the trapped position of the \gls{BEC}.

Atom interferometric imaging is performed by a Ramsey pulse sequence [\cref{fig:setupb}] followed directly by absorption imaging along the $-y$ direction with a \SI{10}{\micro s}-long imaging pulse.
The first \SI{68}{\micro s}-long microwave $\pi/2$ pulse mentioned above creates a coherent superposition of the outcoupled $\ket{1,0}$ and $\ket{2,0}$ states. 
After an evolution time $t_\mathrm{wait}$, a second \SI{68}{\micro s}-long Ramsey pulse is applied to close the interferometer and apply an arbitrary phase shift.
Because of the large hyperfine splitting of the \ce{^{87}Rb} ground state, atoms are imaged spin selectively.

\begin{figure}[btp]
  \centering
  \labelphantom{fig:fig2a}
  \labelphantom{fig:fig2b}
  \labelphantom{fig:fig2c}
  \labelphantom{fig:fig2d}
  \labelphantom{fig:fig2e}
  \labelphantom{fig:fig2f}
  \includegraphics[width=\columnwidth]{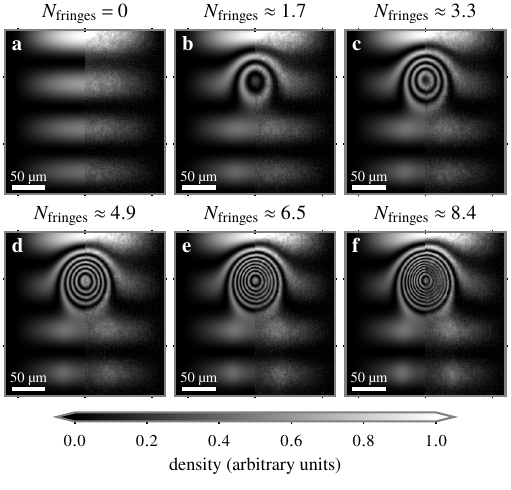}
  \vspace{-1.5\baselineskip}  
  \caption{Atom interferometric imaging of a combined magnetic and optical differential potential for several powers of the dipole potential.
    Each frame is split vertically and shows the theory result (left) compared to the experimental image (right).
    The experimental images were taken with a Ramsey pulse spacing of \SI{0.5}{ms} and are averages of \num{30} experimental runs. The images show the atoms in the $\ket{2,0}$ state.
    The energy differences generated by the dipole beam between the $\ket{2,0}$ and $\ket{1,0}$ are
    \sisetup{round-mode=uncertainty, round-precision=1}%
    \begin{enumerate*}[label=(\alph*)]
    \item \SI[round-mode=none]{0}{\micro K},
    \item \SI{0.16+-0.07}{\micro K},
    \item \SI{0.32+-0.06}{\micro K},
    \item \SI{0.47+-0.04}{\micro K},
    \item \SI{0.6+-0.2}{\micro K}, and
    \item \SI{0.81+-0.07}{\micro K},
    \end{enumerate*}
    and are obtained with uncertainties by least-squares fitting of the experimental data with our model as described in~\EPAPS{}.
    Each title shows the expected number of fringes created by the dipole potential [\cref{eq:fringes}] estimated using the impulse approximation.
  }
  \label{fig:powersequence_with_zero}
  \sisetup{round-mode=none}
\end{figure}

In the presence of a differential potential, the phase evolution between the two Ramsey pulses leads to interference patterns observed in the spin-selective images.
In~\crefrange{fig:setupc}{fig:setupd} and~\cref{fig:powersequence_with_zero}, this potential consists of the magnetic gradient along the vertical direction and [except in \cref{fig:fig2a}] the dipole potential intersecting the atom laser.
The differential potential was constant throughout the experiment, but switched off just before imaging.
The magnetic potential leads to the observation of horizontal interference stripes [\cref{fig:fig2a}], while the dipole potential causes the bull's-eye pattern seen in the upper part of the atom laser.
For short pulse sequences, these approximate contour lines of the differential potential [see~\cref{eq:fringes}].
Phase retrieval~\cite{Bruning:1974, Juarez_Salazar:2018, Schwiegerling:2014} or direct fitting techniques can be used to extract the potential shape, as demonstrated in \cref{fig:setupe}.

In~\cref{fig:powersequence_with_zero}, the depth of the dipole potential is increased from panel to panel, increasing the number of interference rings in the bull's-eye pattern commensurate with the increased phase accumulation between the two Ramsey pulses.
Matching numerics, shown in the left side of each panel in the figure, are in excellent agreement with the experimental images.
Because of the low dipole laser intensities used in \cref{fig:powersequence_with_zero}, the mechanical effects on the atom laser are small in the sense that no pronounced caustics are formed.
For the $\ket{2,0}$ state imaged in \cref{fig:powersequence_with_zero}, the dipole potential is weakly repulsive, leading to a small but increasing suppression of the density in the region below the dipole laser as the power is increased.

\paragraph{Semiclassical analysis and quantitative comparison with experiment}
To theoretically treat our system, we approximate the physics using a semiclassical analysis~\cite{Cartier:2006, Riou:2008, Zapata:2017}.
To start, consider a system uniform along $x$, which is a good approximation near the center of our laser.
Atoms are injected into the system at rest from the trapped cloud at height $z = z_0$ and are immediately subject to a time-dependent potential $V(z, t)$ that includes both gravity and any external potentials, causing the atoms to fall.
Atoms are continuously injected into the system, so atoms imaged at time $t_i$ and height $z_i$ will have been injected at some earlier time $t_0\bigl(t_i, z_i\bigr) \leq t_i$, which must be determined by solving the classical boundary-value problem:
\begin{subequations}
  \begin{gather}
    m\ddot{z}(t) = -\pdiff{}{z}V\bigl(z(t), t\bigr),\\
    \begin{aligned}
      z(t_0) &= z_0, &
      \dot{z}(t_0) &= 0, &
      z(t_i) &= z_i.
    \end{aligned}
  \end{gather}
\end{subequations}
We capture the effects of the various state transitions in the form of the potential $V(z, t)$: if a particle initially in the $\ket{2,0}$ state is subject to Ramsey pulse transition $\ket{2,0} \rightarrow (\ket{2,0} + \ket{1,0})/\sqrt{2}$ at time $t_1$, then we must track two different classical trajectories, having the same potential $V(z, t)$ for $t<t_1$, but different species-dependent potentials $V_{i}(z, t)$ for later times.

In our analysis, we further assume $V(z_0, t) = 0$ for all times, such that the classical Hamiltonian at the injection site $H_0=0$, capturing the essence of coherence in the atom laser: injection occurs resonantly at a fixed energy, keeping the phase of the injected particles constant over time $\psi_0(z_0, t) = \psi_0(z_0)$.

The wave function follows from the path integral:
\begin{subequations}
  \begin{gather}
    \psi(z, t) =
    \int\d{z_0}
    \int \mathcal{D}[q]\; \exp\left\{\frac{\I}{\hbar} S[q]\right\}\psi(z_0, t_0),\\
    S[q] = \int_{t_0}^{t}\!\!\d{t}\left(\frac{m\dot{q}^2}{2} - V\bigl(q(t), t\bigr)\right).
  \end{gather}
\end{subequations}
where the integral is taken over all paths $q(t)$ subject to the boundary conditions $q(t) = z$ and $q(t_0) = z_0$, and $S[q]$ is the classical action.
We assume highly localized injection $\psi(z_0, t_0) \propto \delta(z_0)$, which we take to be about $z_0=0$.  (See~\EPAPS{} for details.)

The \gls{WKB} approximation amounts to expanding the action
\begin{gather}
  S[q+\xi] = S[q] + S'[q]\cdot \xi + \tfrac{1}{2!}S''[q] \cdot \xi\xi + \cdots
\end{gather}
about the classical trajectories $q_\cl$ where $S'[q_\cl] = 0$. 
Keeping only the quadratic fluctuations~\cite{Cartier:2006} with $S[q_\cl]\equiv S(z,t;z_0,t_0)$,
\begin{gather}
  \psi_{\mathrm{WKB}}(z, t) =
  \int\d{z_0}\sqrt{\frac{-\partial^2 S/(2\pi \I \hbar)}{\partial z\partial z_0}}
  e^{\I S / \hbar} \psi(z_0, t_0).
  \exclude{\nonumber\\
  S \equiv S(z,t;z_0,t_0) = \int_{t_0}^{t}\!\!\d{t}\left(
    \frac{m\dot{q}_\cl(t)^2}{2} - V\bigl(q_\cl(t), t\bigr)\right).}      
\end{gather}
If there are multiple trajectories that arrive at the same final position $z(t_i)$ at the time of imaging, one must add these amplitudes to obtain the appropriate interference pattern.

For this position-to-position transition,
\begin{equation}
  \pdiff{S}{z} = p(t),
  \quad \pdiff{S}{z_0} = -p(t_0), 
  \quad\mathrm{and}\quad
  \frac{\partial^2 S}{\partial z\partial z_0} = \pdiff{p}{z_0}.
\end{equation}
If the force is conservative, $E = p^2/2m + V(z) = p_0^2/2m + V(z_0)$ and one recovers
the familiar factor of $\sqrt{p}$ in the denominator of $\psi_{\mathrm{WKB}}(z,t)$:
\begin{align}
  p(z_0, t_0) &= -\sqrt{p_0^2 + 2m\bigl(V(z_0) - V(z)\bigr)}, &
  \pdiff{p}{z_0} &\propto \frac{1}{p}.
\end{align}
The semiclassical problem is thus reduced to solving for the classical trajectories of particles injected at $q(t_0) = (x_0, z_0)$ that end up at $q(t_i) = (x, z) \equiv \vect{x}$ in the image.

When preparing numerical simulations for the experiment, we interfere two different trajectories: those of the particles which remain in state $\ket{2, 0}$ ($\psi_1$) and those which start in the state $\ket{2, 0}$ but are converted to state $\ket{1,0}$ for times between the two Ramsey pulses ($\psi_2$).
This procedure has a few deficiencies.
First, the semiclassical amplitudes diverge at the turning point $z=z_0$.
This can be remedied by using Airy functions, but to demonstrate the accuracy of the pure semiclassical calculation, we simply exclude the region close to the injection site in our comparisons.
Second, the model assumes instantaneous state transitions.
We mitigate this by allowing the transition to occur at a time slightly shifted from the middle of the transition window that accounts for the acceleration of the particles.
A proper semiclassical accounting for this effect requires a multicomponent \gls{WKB} approximation~\cite{Dijk:1979, Littlejohn:1991, Emmrich:1996, Hagino:2004}, which is much more complicated and not needed here.

The interference pattern $I_n \propto \abs{\psi_{1} + \psi_2}^2$ can be modeled as $I_n(\vect{x}) \approx a(\vect{x}) + b(\vect{x})\cos\bigl(\phi(\vect{x}) + \theta_n\bigr)$ where $\theta_n$ is an experimentally controllable phase.
Phase retrieval techniques~\cite{Bruning:1974, Juarez_Salazar:2018, Schwiegerling:2014} can efficiently extract the difference in actions between the paths $\hbar \phi(\vect{x}) = S_{1}(\vect{x}) - S_{2}(\vect{x}) + \text{const.}$, from which the differential potential can be extracted.
This requires imaging at least three different values of $\theta_n$ (see~\EPAPS{}).

Alternatively, if the form of the potential is known up to a few parameters, then the interference pattern can be directly modeled from a single value of $\theta_n$, allowing high-precision fitting of these parameters.
This analysis is the basis for our numerical simulations and leads to a quantitative explanation of the experimental data, as demonstrated in \cref{fig:powersequence_with_zero}.
With a few simplifying approximations that we call the ``impulse approximation'' -- assuming weak potentials do not appreciably deflect the particles, and the transitions and $t_{\text{wait}}$ are sufficiently fast such that the particles do not fall significantly during the Ramsey pulse sequence -- one obtains the following density pattern, and corresponding expected number of maxima (fringes) in the interference pattern:
\begin{align}
  n^{\text{Ramsey}}_{\ua} &\propto 1 - \cos\Bigl(\frac{t_{\mathrm{wait}}}{\hbar}\delta V\Bigr),
  &
    \label{eq:fringes}
  N_{\mathrm{fringes}} \approx \frac{t_{\mathrm{wait}}\;\delta V_{\max}}{2\pi \hbar},
\end{align}
where $\delta V_{\max}$ is the maximum of the differential potential $\delta V(\vect{x})$,  (see also \EPAPS{}).
This agrees well with the full calculations and experiments, as shown in \cref{fig:powersequence_with_zero}.

\paragraph{Magnetic field mapping}
So far, we have demonstrated the effect of combined magnetic and optical differential potentials using Ramsey pulse sequences between the $\ket{2,0}$ and $\ket{1,0}$ states.  This transition is only weakly sensitive to magnetic fields due to second order Zeeman effects.
In applications where a greater sensitivity to magnetic fields is desired, a strongly magnetic field dependent transition such as the one between the stretched states $\ket{1,-1}$ and $\ket{2,-2}$ can be employed. 
In \ce{^{87}Rb}, this transition shifts by \SI{-2.1}{MHz/G} in low fields, compared to \SI{11}{kHz/G} for the $\ket{1,0}$ to $\ket{2,0}$ transition in a bias field of \SI{10}{G}. 
When using magnetically sensitive transitions, care must be taken that the Ramsey pulses affect the entire atom laser, otherwise a state transfer will occur in only a small region, which can be used for fluid flow tracing~\cite{Mossman:2021}. 

\begin{figure}
  \centering
  \labelphantom{fig:fig3a}
  \labelphantom{fig:fig3b}
  \labelphantom{fig:fig3c}
  \labelphantom{fig:fig3d}
  \includegraphics[width=0.45\textwidth]{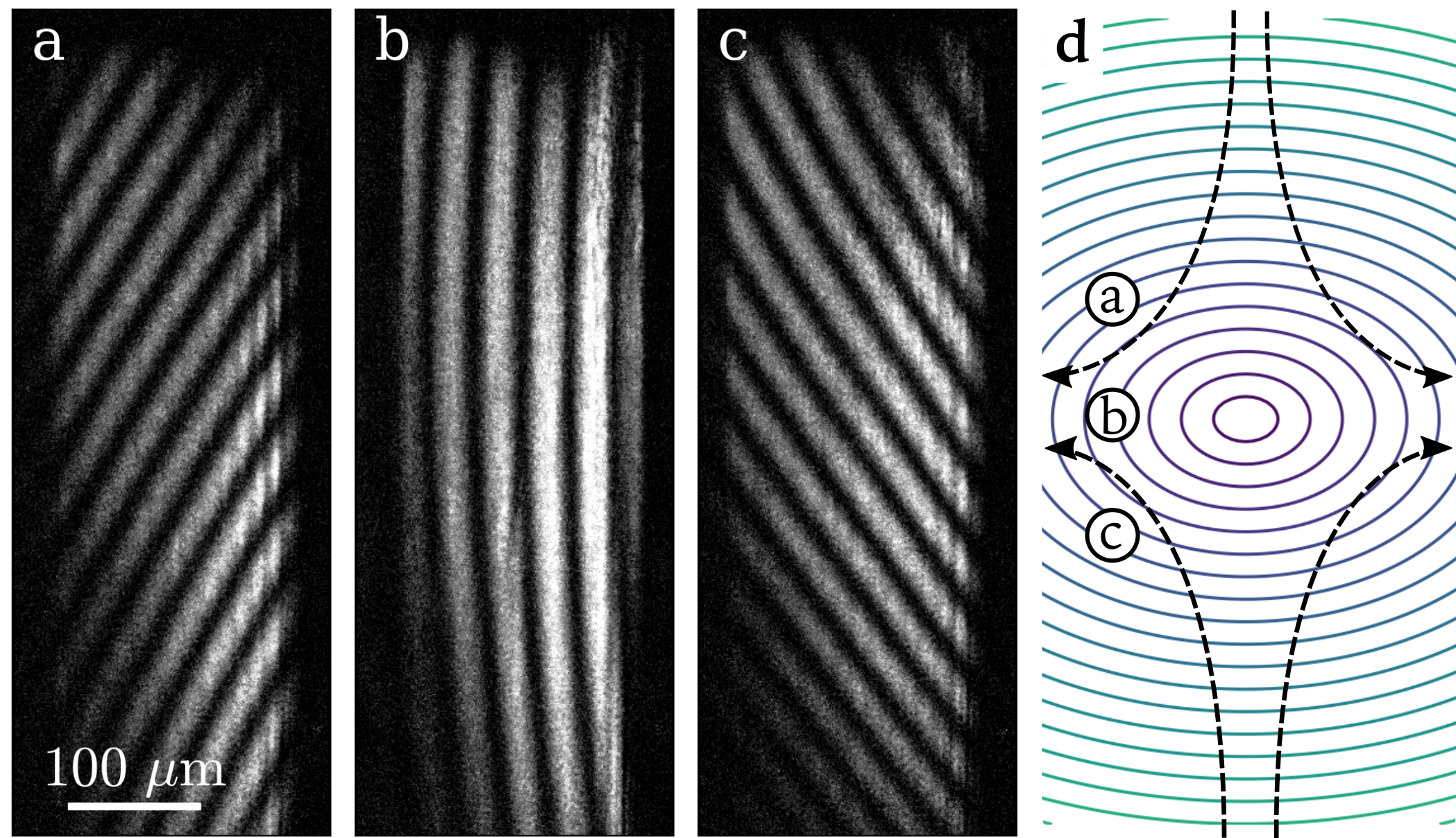}
  \caption{Atom interferometric fringe pattern using the magnetically sensitive transition between $\ket{1,-1}$ and $\ket{2,-2}$.
    A magnetic quadrupole field with gradient $\d{B}/\d{z}=\SI{140\pm10}{mG/cm}$ was placed with its center to the right and
    \begin{enumerate*}
    \item\label{item:3a}slightly below, 
    \item\label{item:3b}right next to, and 
    \item\label{item:3c}slightly above the atom laser.
      The Ramsey pulses are spaced by \SI{2}{ms}.
      The images show the atoms detected in the $\ket{2,-2}$ state after the Ramsey sequence.
      Each image is from a single repetition of the experiment. 
    \item\label{item:3d}Schematic representation of quadrupole field (not to scale).
      Letters correspond to \crefrange{item:3a}{item:3c} and indicate the position of quadrupole field.
      The magnetic gradient has been calculated from the interference pattern in \cref{item:3b}.
    \end{enumerate*}
  }  
  \label{fig:zebra1m12m2}
\end{figure}

We demonstrate the capability of using the $\ket{1,-1}$ to $\ket{2,-2}$ transition to detect small magnetic gradient fields in \cref{fig:zebra1m12m2}. 
Here, an atom laser is generated from a \gls{BEC} confined in a purely optical trap: a large-diameter dipole beam is employed to provide mostly radial confinement, and two repulsive, thin dipole sheets are added as ``end caps'' on the left and right side of the \gls{BEC} to provide axial confinement. 
The atom laser is then realized by ramping down the intensity of the large-diameter dipole beam to create a wide atom laser without relying on any preexisting magnetic gradient for the output coupling. 
To generate a test pattern, a magnetic quadrupole field with an axial gradient of \SI{140\pm10}{mG/cm} was added, approximately 2 orders of magnitude smaller than the gradient used for the previous images. The position of the quadrupole field zero was adjusted relative to the imaging window using small magnetic bias fields, as shown schematically in \cref{fig:fig3d}.  The results clearly show the tilt of the equipotential lines in the magnetic quadrupole field, demonstrating the capability of imaging magnetic field gradients in a single experimental run. 
 
\begin{figure}
  \centering
  \labelphantom{fig:spinechoa}
  \labelphantom{fig:spinechob}
  \includegraphics[width=0.45\textwidth]{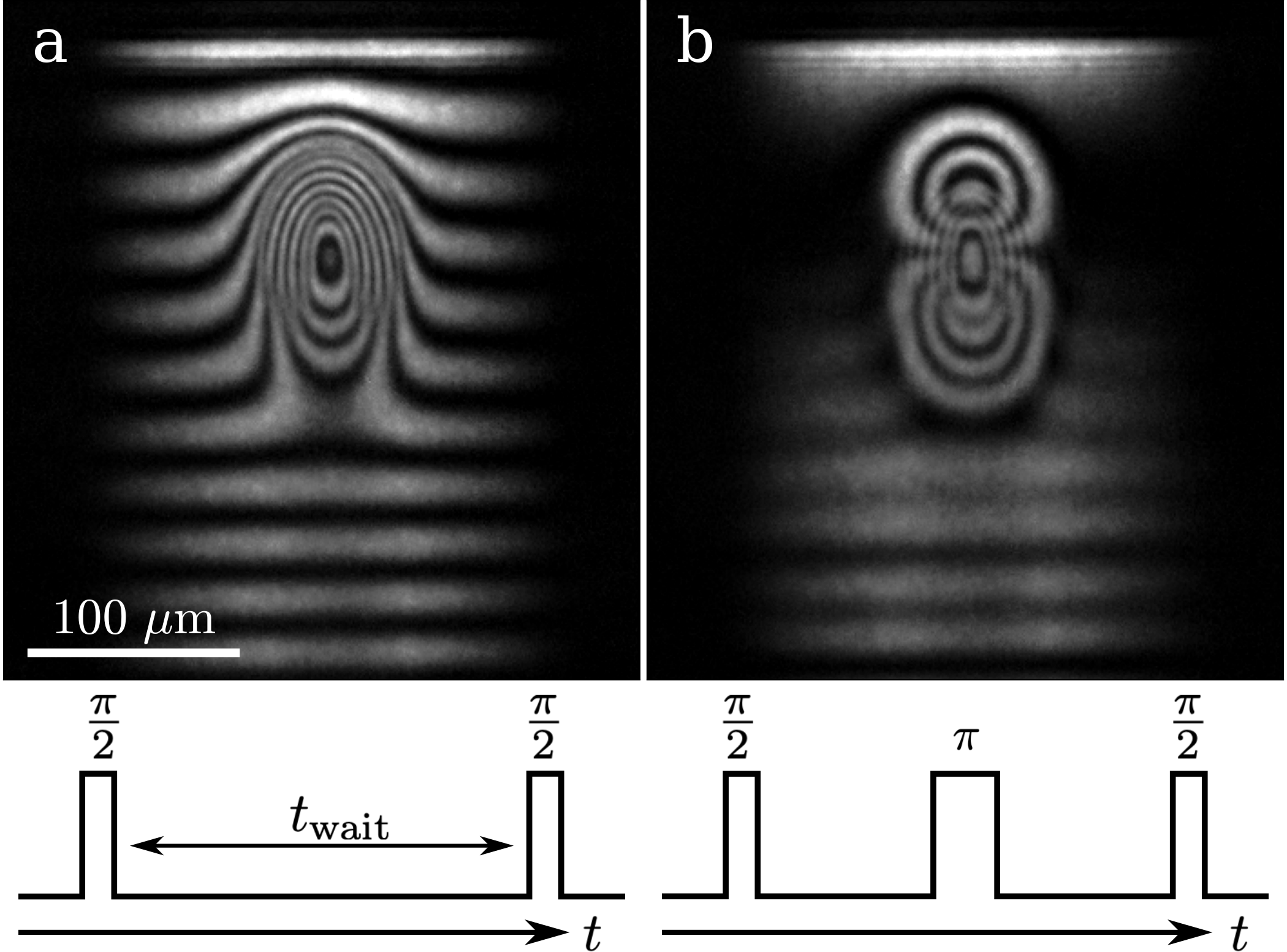}
  \caption{Comparison of
    \begin{enumerate*}[label=(\alph*)]
    \item Ramsey imaging and
    \item spin-echo imaging.
    \end{enumerate*}
    Pulse sequence timelines are depicted below each image (not drawn to scale). 
    Both images show the $\ket{2,0}$ state \SI{0.5}{ms} after the end of the pulse sequence. 
    Total wait time between the $\pi/2$ pulses is $t_\mathrm{wait}=\SI{1504}{\micro s}$ in both cases. 
    Images have been averaged over \num{30} independent experimental runs of the same parameters.}
  \label{fig:spinecho}
\end{figure}

\paragraph{Spin-echo imaging}
Atom interferometric techniques provide great flexibility for the design of experimental sequences. 
While the experiments described above have all used a Ramsey pulse sequence, extended sequences can be employed to enhance specific features. 
One example is demonstrated in \cref{fig:spinecho} where the Ramsey sequence has been augmented by inserting an additional $\pi$ pulse, realizing a spin-echo sequence. 
Such a sequence can be used to cancel the effects of constant differential potentials and to produce contour lines of the gradient along the direction of motion.
For \cref{fig:spinechoa}, a pulse spacing $t_\mathrm{wait}$ of \SI{1504}{\micro s} between the two pulses of a Ramsey sequence was used, with each $\pi/2$ pulse lasting \SI{53}{\micro s} and connecting the $\ket{2,0}$ and $\ket{1,0}$ states. 
The corresponding spin-echo sequence shown in \cref{fig:spinechob} was chosen to have the same total length of the sequence between the Ramsey pulses.
The two panels in \cref{fig:spinecho} show that, over a significant region in the upper part of the atom laser, the spin-echo sequence suppresses the horizontal stripes caused by the weak magnetic gradient, while imaging the gradient of the dipole potential along the vertical direction. 
The oval-shaped features in the center of \cref{fig:spinechob} are a consequence of an inefficiency of the $\pi$ pulse in the center of the potential where light shifts are significant. 
Similarly, the cancellation of the horizontal stripe pattern near the bottom of the image is incomplete as the $\pi$ pulse is slightly shifted out of resonance here. 
In principle, these issues could be mitigated if sufficient microwave power is available by applying shorter pulses with larger linewidths. 
This demonstrates the capabilities of suitably chosen pulse sequences to enhance or modify the imaging contrast of desired features.

\paragraph{Conclusion}
As demonstrated in this work, atom interferometric imaging with an atom laser is a powerful tool for detecting and evaluating differential potentials over a large two-dimensional area. 
With the availability of highly tunable pulse sequences and several accessible spin states for measuring a variety of differential potentials, this technique is highly versatile and can be adapted for a wide set of applications. 
As an applied example, relating to work presented in Ref.~\cite{Taylor:2021}, one can consider applications to material science where a material under study is placed parallel to the sheet of an atom laser, detecting the magnetic fields emanating from the material by imaging them in the plane of the atom laser using the interferometric technique.

\begin{acknowledgments}
  P.E.\ acknowledges \supportfromNSFgrant[PHY]{1912540} and from the Ralph G. Yount Distinguished Professorship at \fundrefWSU.
  M.E.M.\ acknowledges \supportfromNSFgrant[PHY]{2137848} and from the Clare Boothe Luce Professorship Program of the Henry Luce Foundation.
  R.A.C.\ and M.M.F.\ acknowledge \supportfromNSFgrant[PHY]{2012190}.
\end{acknowledgments}

\exclude{
\paragraph{Author Contributions}
M.E.M. and P.E. conceived the experiment and performed experiments and data analysis.
R.A.C and M.M.F. performed theoretical calculations and numerical simulations.
All authors discussed the results and contributed to the writing of the manuscript.

\paragraph{Competing Interests}
The authors declare no competing interests.

\paragraph{Materials \& Correspondence}
Please direct any questions or requests concerning this article to P.~Engels or M.~M.~Forbes.

\paragraph{Code Availability}
All relevant code used for numerical studies in this work is available from the corresponding authors upon reasonable request.
Additional code for visualizing 3\textsc{d} caustics is available from~\cite{gitlab:catastrophe_atom_optics}.

\paragraph{Data Availability}
All relevant experimental and numerical data sets in this work will be made available from the corresponding authors upon reasonable request.
}

\section{Supplementary Material}
\label{sec:supp}

\subsection*{Differential dipole potential}
The differential dipole potential is generated with a laser locked to the $F=3$ to $F'=4$ transition of \ce{^{85}Rb}. 
The light is $\pi$-polarized with respect to the bias field applied to the atoms. 
With laser powers on the order of a hundred nW sent to the atoms and a Gaussian beam waist of $39~\mu$m, differential potential depths of few hundreds of nK are generated at the atom laser sheet. 
Potential depths can be theoretically estimated, for example by following Ref.~\cite{Deissler_2008}. 

\subsection*{Output Ports}
\begin{figure}[htb!]
  \centering
  \includegraphics[width=\linewidth]{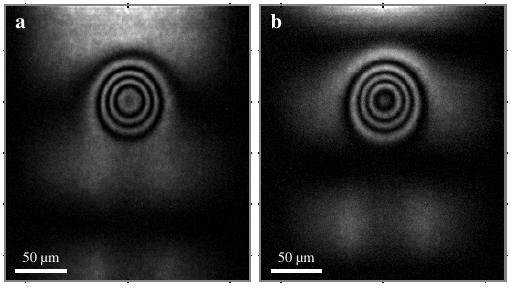}
  \vspace{-1.75\baselineskip}   
  \caption{The
    \begin{enumerate*}
    \item $\ket{1, 0}$ and
    \item $\ket{2, 0}$ states form two complementary output ports of the interferometer that can separately be imaged.
    \end{enumerate*}
    The images shown here are taken with a Ramsey pulse spacing of \SI{0.25}{ms}.
    Each image has been averaged over \num{30} runs of the experiment.
    As the imaging transition is stronger for the $\ket{2,0}$ state, the intensities of the two panels have been independently scaled.
  }
  \label{fig:output}
\end{figure}

\Cref{fig:output} shows the two complementary output ports of the interferometer, imaging the $\ket{1,0}$ or $\ket{2,0}$ states after the Ramsey sequence, respectively.
This provides an alternative way to obtain the information shown in \Crefrange{fig:setupc}{fig:setupd} for two phase shifts separated by a $\pi$ phase shift.

\subsection*{Influence of Ramsey time}
\begin{figure}
    \centering
    \labelphantom{fig:fig5a}
    \labelphantom{fig:fig5b}
    \labelphantom{fig:fig5c}
    \labelphantom{fig:fig5d}
    \labelphantom{fig:fig5e}
    \labelphantom{fig:fig5f}
    \includegraphics[width=\linewidth]{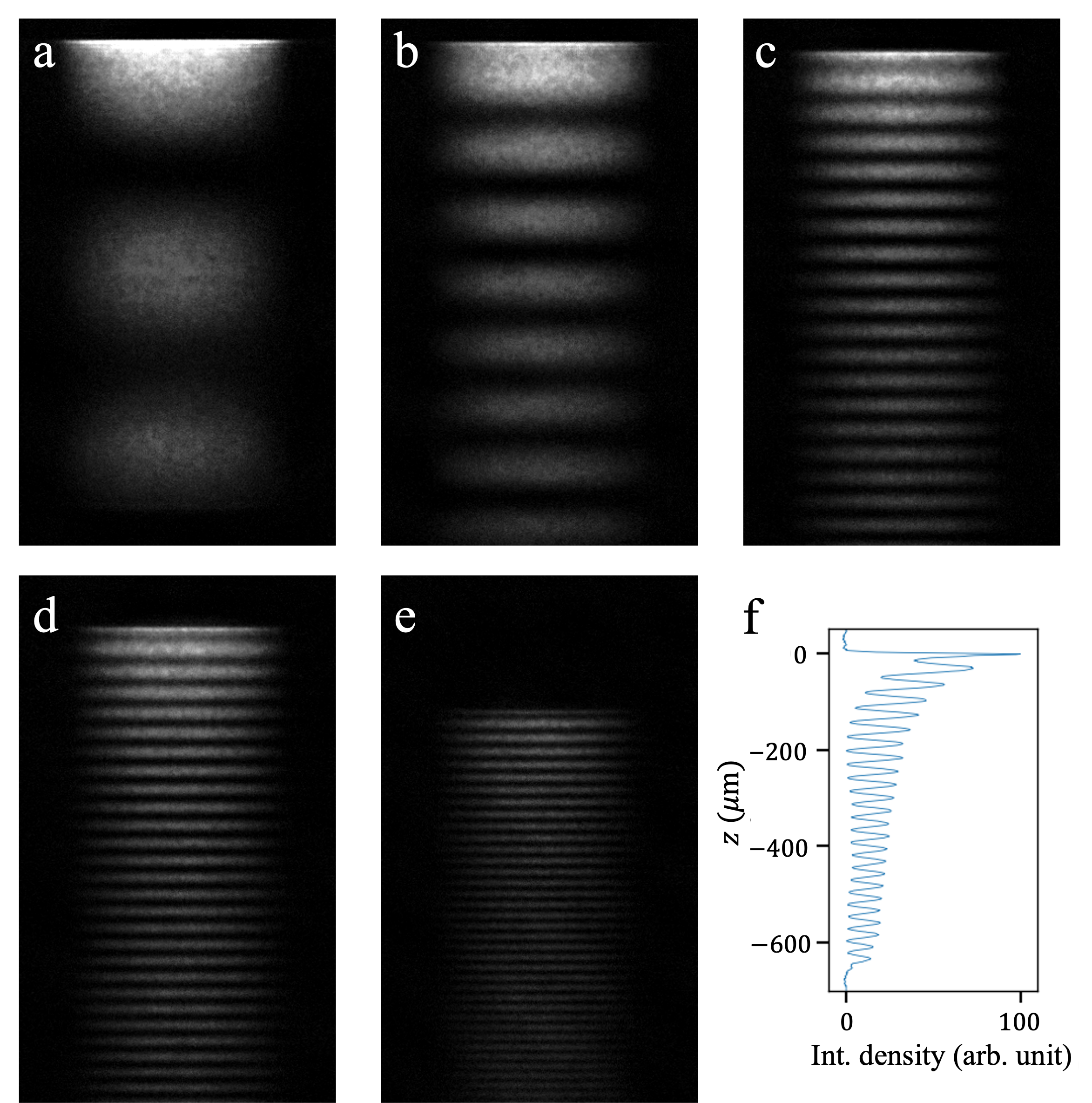}
    \caption{Atom interferometric fringe pattern as a function of Ramsey pulse spacing.
      Images are taken in the presence of a magnetic gradient; no dipole potential has been applied.
      The Ramsey pulse spacing is (a) \SI{0.1}{ms}, (b) \SI{0.5}{ms}, (c) \SI{1.5}{ms}, (d) \SI{2.5}{ms}, (e) \SI{5}{ms}.
      Each of the two Ramsey pulses has a pulse length of \SI{68}{\micro s}.
      Panel (f) shows the horizontally integrated cross section of (c).
      All image intensities are scaled with the same factor for comparison.
      Each panel has been averaged over 10 independent runs with the same experimental parameters.}
    \label{fig:zebrastudies}
\end{figure}

The observed spacing between the interference fringes is a function of the differential phase accumulation occurring between the two Ramsey pulses. 
This phase accumulation not only depends on the depth of the potential, but also on the chosen time between the two pulses. 
This dependence on the pulse spacing is demonstrated in \cref{fig:zebrastudies}, where a magnetic gradient of \SI{-25.1}{G/cm} was used to generate a differential potential for the $\ket{2,0}$ and $\ket{1,0}$ state. 
Here, \crefrange{fig:fig5a}{fig:fig5e} were taken under identical conditions but for Ramsey pulse spacings of $t_\mathrm{wait}= \{0.1,0.5,1.5,2.5,5.0\}~$ms, respectively. 
The density of lines caused by the magnetic gradient increases with the pulse spacing as expected. 
\cref{fig:fig5f} shows a horizontally integrated cross section of \cref{fig:fig5c}. 

\subsection*{Determination of the magnetic gradient}

\begin{figure}[htbp]
     \centering
     \includegraphics[width=\linewidth]{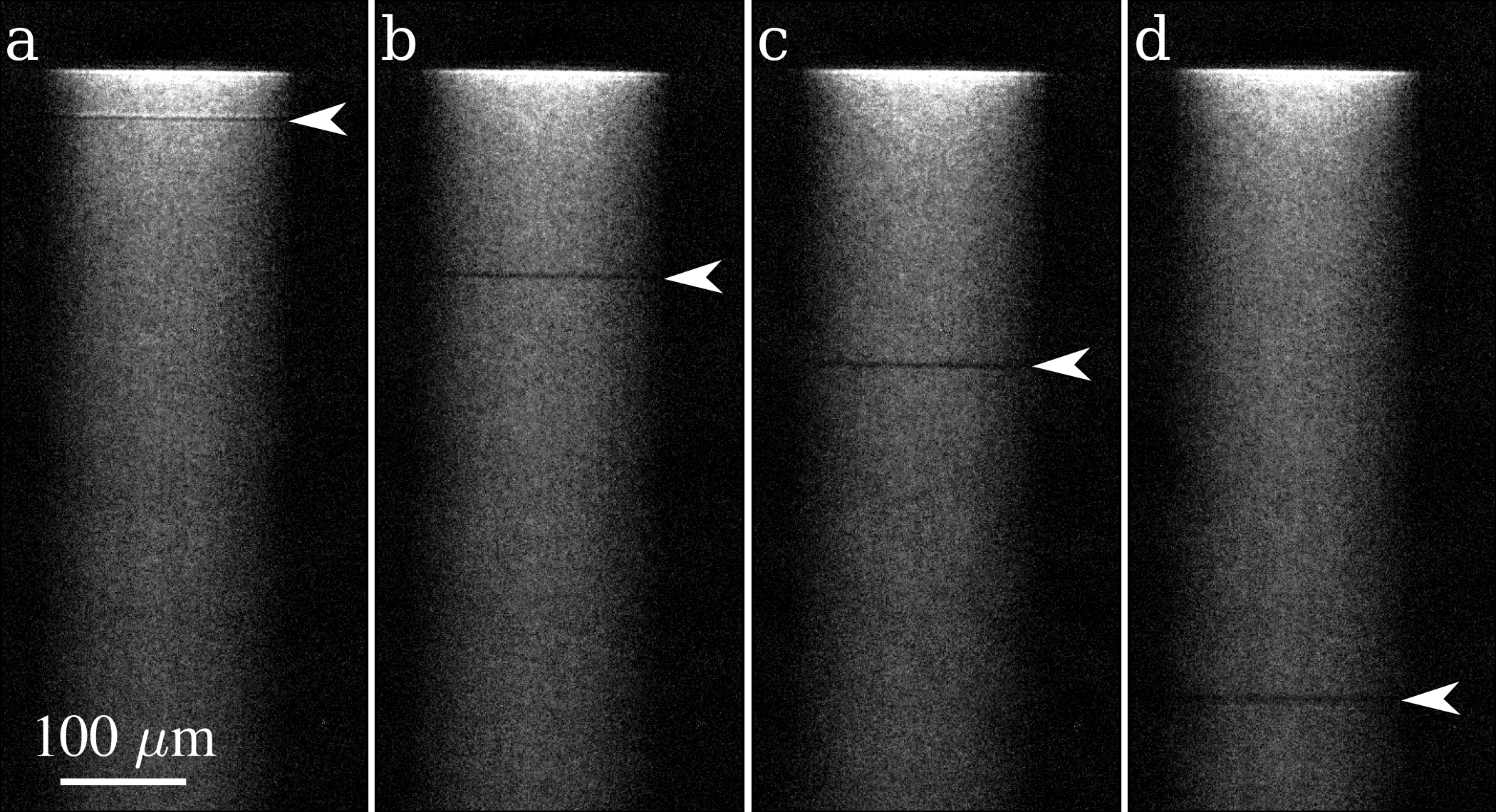}
     \caption{Spectroscopic mapping of magnetic field contour lines using fluid flow tracing.
       Microwave frequency \textbf{a} \SI{6.8147}{GHz}, \textbf{b} \SI{6.814}{GHz}, \textbf{c} \SI{6.8136}{GHz}, and \textbf{d} \SI{6.8121}{GHz}.
       The white markers indicate the position where atoms are transferred.
       The intensity for all images has been scaled identically for comparison.
       Each panel is the result of a single run of the experiment.}
     \label{fig:bfieldtracers}
 \end{figure}

To determine the magnetic gradient present for the data shown in \cref{fig:setup}, \cref{fig:powersequence_with_zero}, \cref{fig:spinecho} and \cref{fig:zebrastudies}, a spectroscopic technique is employed, where we use the highly magnetic sensitive transition between the $\ket{2,-2}$ and $\ket{1,-1}$ states. 
An atom laser is generated comprised of atoms in the $\ket{2,-2}$ hyperfine state. 
Using a weak and brief microwave pulse with a duration of \SI{50}{\micro s}, a thin, horizontal stripe of atoms is transferred into the $\ket{1,-1}$ state immediately before an image is taken. 
In contrast to the images shown in the main text where the Ramsey pulses transferred the whole atom laser, here a microwave pulse only transfers a small stripe of atoms.
In images taken of the $\ket{2,-2}$ atom laser, this stripe appears as a dark horizontal line as indicated by the white markers in \cref{fig:bfieldtracers}. 
The position of this line is determined, and the corresponding magnetic field is calculated based on the applied microwave frequency and the Breit-Rabi formula. 
This procedure is repeated for a variety of microwave frequencies, resulting in a map of the magnetic field as a function of position. 

\subsection*{Fitting the Data}

Simulations are fit to the experimental data by varying important system parameters and numerically minimizing an overlap function that yields a measure of $\chi^2_r$ for the simulation.
We identify as meaningful the following parameters:
\begin{itemize}
\item $V_0$: The potential depth generated by the Gaussian beam, proportional to the laser power.
  A higher power leads to more interference fringes in the optical region, and to more pronounced shadows below the beam.
  In a typical experiment, one controls the power of the laser, hence $V_0$ depends on both the power and the waist of the optical beam: a wider beam distributes the power over a large area, reducing $V_0$.
    Thus, for our simulations, we use $V_0$ which directly correlates with the number of fringes according to~\cref{eq:fringes} if the impulse approximation is valid.
\item $w$: The waist of the optical beam.
  A larger waist shifts the apparent center of the potential down slightly.
\item $z_0$: The location of the center of the optical beam in the z direction, relative to the atom laser injection site.
  This is a convenience parameter for aligning the simulations to the experimental data and does not affect the major physical features.
\item $t_\mathrm{wait}$: The time between Ramsey pulses entirely dictates the horizontal interference pattern in the atom laser.
  The wait time also affects the number of interference fringes within the region of the optical laser, with longer wait times leading to more interference fringes (see~\cref{eq:fringes}).
\item $R_\mathrm{ab}$: The relative potential depths for particles occupying the $\ket{2,0}$ and $\ket{1,0}$ states.
\item $V_\text{dev}$: This is a linear piece in the form of the optical potential, modelling a deviation in the potential from a pure Gaussian. This affects the optical fringes by making the beam more powerful, and also slightly shifts the apparent location of the center of the beam.  This parameter is intended to model a tiny but visible asymmetry in the fringe pattern above and below the center of the beam.
\item $\lambda_i$: an offset for the time at which the particles begin to feel the $\pi/2$ pulse. 
This parameter is implemented in order to simulate that particles in the atom laser do not instantly feel the effects of the $\pi/2$ pulse, but will on average feel it at the center of its pulse time. This is well approximated by a value of $\lambda_i=0.5$.
\end{itemize}

\begin{table}[htbp]
  \sisetup{
    detect-weight = true,
    detect-all = true,
    round-mode = uncertainty,
    round-precision = 1,
    table-number-alignment = center,
    table-format = 2.3,
  } %
  \begin{tabular}{l
    S[table-format=3.6]                                          
    S[table-format=3.5]                                          
    S
    S
    S[table-format=3.5]                                          
    S[round-mode = figures, round-precision=2, table-format=1.1] 
    }
    \toprule
      & {$V_0$ [\si{\micro K}]} & {$z_0$ [\si{\micro m}]} & {$w$ [\si{\micro m}]} & {$R_{\text{ab}}$} & {$V_{\text{dev}}$} & {$\chi^2_r$}\\
    \midrule
    b) & 0.164497403810694+-0.06722372965231025& 89.7024007066054+-0.93691476075139& 38.63057029750185+-2.4004819035284313& 2.644673098065572+-4.557056073806767& -0.0012743522842987229+-0.0899374839214449 & 3.0464032764742113 \\
    c) & 0.3165470809653781+-0.05775751699055267& 90.01653507618647+-0.6628248907397924& 38.635713882077226+-1.3561799213155756& 1.4931224348823067+-2.1768983897421648& 0.002031465750442934+-0.0392103056170217 & 2.7679496120025515 \\
    d) & 0.47354342564340334+-0.0381412871576027& 89.98833388366823+-0.7142360822845816& 39.0079437127573+-0.8449473490884898& 0.9510242969681479+-1.2896464583298188& 0.0020194239458994918+-0.0158600203962235 & 3.7429489592916925 \\
    e) & 0.6264513882056654+-0.19466601651502017& 90.00597130019858+-1.297283587431257& 38.88084083366674+-0.6887499554574471& 0.7137396046912676+-1.1494216090208154& -0.0010905516671286663+-0.06873664824729726 & 5.281240561504619 \\
    f) & 0.8068527443765987+-0.07325775395723865& 90.0065851016233+-0.4999634092086377& 39.141826838109324+-0.5784414507443083& 0.6715022190421656+-0.721390853799493& 0.0021297368003892264+-0.019724144030927534 & 9.17642911278388 \\
      \bottomrule
  \end{tabular}
  \exclude{
  \begin{tabular}{l
    S[table-format=3.6]                                          
    S[table-format=3.5]                                          
    S
    S[round-mode = figures, round-precision=2, table-format=1.1] 
    }
    \toprule
      & {$V_0$ [\si{mK}]} & {$z_0$ [\si{\micro m}]} & {$w$ [\si{\micro m}]} & {$\chi^2_r$}\\
    \midrule
    b) & 0.164497403810694+-0.06722372965231025& 89.7024007066054+-0.93691476075139& 38.63057029750185+-2.4004819035284313 & 3.0464032764742113 \\
    c) & 0.3165470809653781+-0.05775751699055267& 90.01653507618647+-0.6628248907397924& 38.635713882077226+-1.3561799213155756 & 2.7679496120025515 \\
    d) & 0.47354342564340334+-0.0381412871576027& 89.98833388366823+-0.7142360822845816& 39.0079437127573+-0.8449473490884898 & 3.7429489592916925 \\
    e) & 0.6264513882056654+-0.19466601651502017& 90.00597130019858+-1.297283587431257& 38.88084083366674+-0.6887499554574471 & 5.281240561504619 \\
    f) & 0.8068527443765987+-0.07325775395723865& 90.0065851016233+-0.4999634092086377& 39.141826838109324+-0.5784414507443083 & 9.17642911278388 \\
      \bottomrule
  \end{tabular}
}
  \sisetup{round-mode=none}
  \caption{Best fit parameter values for our model for each of the data-sets in \crefrange{fig:fig2b}{fig:fig2f}.
    (\cref{fig:fig2a} has no information about the differential-dipole potential, so we do not fit it.)
    The error estimates are the 1$\sigma$ deviations based on a standard minimization of the reduced $\chi^2_r$ (\cref{eq:chi2r}), scaling the final results so that $\chi^2_r = 1$.
    The fits include $\num{73700} = 268\times 275$ pixels for $\abs{x} < \SI{120}{\micro m}$ and $z \in [\SI{-275}{\micro m}, \SI{-30}{\micro m}]$.
    This excludes the original cloud and any systematic deviations from neglecting the Airy functions in our semiclassical calculation as well as any effects from non-uniform fields below the injection site.
   }
   \label{tab:fits}
\end{table}

 The horizontal striping effect is purely due to the Ramsey pulse spacing $\text{t}_{\text{wait}}$, while the fringes in the vicinity of the dipole potential are due to both $\text{t}_{\text{wait}}$ and $V_0$.

We perform a least-squares minimization over these parameters of
\begin{gather}
  \label{eq:chi2r}
  \chi^2_r = \frac{1}{\nu}\sum
  \frac{(n_{\text{theory}} - n_{\text{experiment}})^2}{(\sigma n_{\max})^2}
\end{gather}
where we estimate 
$\sigma \approx \num{0.00056}$ 
as the standard deviation of the background thermal fluctuations taken from a dark patch of the image, scaled by the normalization factor applied to the data before minimization.
The sum is over $\num{73700} = 268\times 275$ pixels for $\abs{x} < \SI{120}{\micro m}$ and $z \in [\SI{-275}{\micro m}, \SI{-30}{\micro m}]$.
Fitting 5 parameters gives $\nu = \num{73694}$ degrees of freedom.

\begin{figure}[htbp]
  \centering
  \labelphantom{fig:S2a}
  \labelphantom{fig:S2b}
  \labelphantom{fig:S2c}
  \labelphantom{fig:S2d}
  \labelphantom{fig:S2e}
  \labelphantom{fig:S2f}
  \includegraphics[width=\columnwidth]{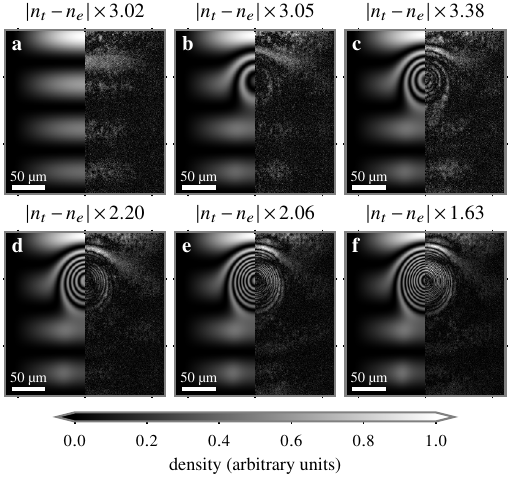}
  \caption{Residuals of the best fit semiclassical model with the experimental data.
    The frames correspond to those in \cref{fig:powersequence_with_zero}.
    Each frame is split vertically and shows the theory result $n_t \equiv n_{\mathrm{theory}}$ (left) compared to the scaled residuals (right).
    The residuals are scaled so that the maximum residual matches the maximum theory density to facilitate comparison.
    These scaling factors are listed in the title of each frame.
  }
  \label{fig:S2}
\end{figure}

The final $\chi^2_r$ values listed in \cref{tab:fits} indicate that the fits are not consistent with this thermal noise.
The residuals shown in \cref{fig:S2} give some hints about deficiencies of the model.
A large residual in the troughs of the interference fringes are likely due to several effects include:
\begin{itemize}
\item Using a single injection point.  A more accurate model would convolve the results over the full injection region, reducing the purity of the interference between the states.
\item Each experimental image is an average of $\sim 30$ shots.
  The clarity of the interference patterns demonstrates the repeatability of experiment, but one expects slight variations in environmental conditions between images that could shift parameters like $z_{d}$ slightly from run to run, smearing the experimental images most noticeably in the dark regions.
  One might mitigate this by individually fitting each shot.
\item Our model assumes perfectly Gaussian beams.
  There are likely deviations beyond what we tried to model through $V_{\mathrm{dev}}$ and $R_{\mathrm{ab}}$ due to optics and 3D geometry.
  An interesting future direction is to try to perform parameter-free reconstructions of the potential from the interference patterns, but for accuracy, using a well-defined form is always preferable.
\end{itemize}

\subsection*{Coherence of the Atom Laser}
As a check of our semi-classical analysis, we model a pure atom laser in a constant
gravitational field through the two-component Hamiltonian (see also~\cite{Riou:2008,
  Kramer:2006, Kramer:2006err, Harkonen:2010} for similar analysis):
\begin{gather*}
  \I\hbar \pdiff{}{t}\ket{\Psi} = \op{\mat{H}}\ket{\Psi}, \qquad
  \ket{\Psi(t)} = \begin{pmatrix}
                 \ket{\psi_a(t)}\\
                 \ket{\psi_b(t)}
  \end{pmatrix}, \\
  \op{\mat{H}} = \begin{pmatrix}
    \frac{\op{p}_z^2}{2m} + mg\op{z} & \Omega e^{\I\omega t}\\
    \Omega e^{-\I\omega t} & \frac{\op{p}_z^2}{2m} + V(\op{z})
  \end{pmatrix}.
\end{gather*}
The idea of an atom laser is that a large reservoir of the lower component is held in the trapping potential $V(z)$.
The off-diagonal coupling converts this lower component $\ket{\psi_b}$ to the upper component $\ket{\psi_a}$,
\footnote{In our experiment $\ket{\psi_b}$ corresponds to the hyperfine state $\ket{1, -1}$, while $\ket{\psi_a}$ initially corresponds to the hyperfine state $\ket{2,0}$. At time $t\approx t_1$ the $\ket{2,0}$ state is mixed with the $\ket{1,0}$ hyperfine state, which would require a three-component formalism.} 
which then falls in the gravitational field.
If the off-diagonal coupling is small, then one can treat the trapped component as a constant, and we have the following coupled equation (essentially neglecting the lower-left block):
\begin{align*}
  \I\hbar \ket{\dot{\psi}_a(t)} 
    &= \left(\frac{\op{p}_z^2}{2m} + mg\op{z}\right)\ket{\psi_a(t)}
       + \Omega e^{\I\omega t}\ket{\psi_b(t)}\\
   \I\hbar \ket{\dot{\psi}_b(t)} 
   &= \left(\frac{\op{p}_z^2}{2m} + V(\op{z}) - E_b\right)\ket{\psi_b(t)}.
\end{align*}
After some time, the states become quasi-stationary, and we can write:
\begin{gather*}
  \ket{\psi_b(t)} = \ket{\psi_b}e^{-\I E_b t/\hbar}, \qquad
  \ket{\psi_a(t)} = \ket{\psi_a}e^{\I (\hbar \omega - E_b) t /\hbar},\\
  \begin{aligned}
    \left(\frac{\op{p}_z^2}{2m} +  mg\op{z} - \I\hbar \partial_t\right)\ket{\psi_a(t)}
    &= - \Omega e^{\I(\hbar \omega - E_b)t/\hbar}\ket{\psi_b},\\
   \left(\frac{\op{p}_z^2}{2m} + V(\op{z}) - E_b\right)\ket{\psi_b} &= 0,\\
    \left(\frac{\op{p}_z^2}{2m} + mg\op{z} + \hbar \omega - E_b\right)\ket{\psi_a}
    &= -\Omega \ket{\psi_b},\\
    \left(\frac{\op{p}_z^2}{2m} + mg(\op{z} + z_0)\right)\ket{\psi_a}
    &= -\Omega \ket{\psi_b}.\\
  \end{aligned}
\end{gather*}
The last equation is simplified by setting $mgz_0 = \hbar \omega - E_b$, which we can redefine as the zero of our coordinate system $z \rightarrow z - z_0$.
Doing this, and rescaling $\tilde{z} = z/\xi$, we have:
\begin{gather*}
  \left(-\pdiff[2]{}{\tilde{z}} + \tilde{z}\right)\psi_a(\tilde{z}) = -\frac{2m\Omega \xi^2}{\hbar^2}\psi_b(\tilde{z}),\qquad
  \xi = \sqrt[3]{\frac{\hbar^2}{2m^2g}}.
\end{gather*}

The homogeneous solution can be expressed in terms of the Airy functions:
\begin{gather*}
  \psi(\tilde{z}) = a\Ai(\tilde{z}) + b\Bi(\tilde{z}),
\end{gather*}
where $y=\Ai(\tilde{z})$ and $y=\Bi(\tilde{z})$ are the orthogonal real solutions with $\lim_{\tilde{z}\rightarrow\infty}\Ai(\tilde{z}) = 0$ to
\begin{gather*}
  y'' = \tilde{z}y,
\end{gather*}
and the full solution can be expressed in terms of the Green's function (see e.g.~\cite{Bracher:1998, Vallee:2010})
\begin{align*}
  G(\tilde{z},\tilde{z}') &= -\pi \begin{cases}
    \Ai(\tilde{z}')\Ci(\tilde{z}) & \tilde{z} \leq \tilde{z}',\\
    \Ai(\tilde{z})\Ci(\tilde{z}') & \tilde{z} \geq \tilde{z}'
  \end{cases}, \\
  \Ci(\tilde{z}) &= \Bi(\tilde{z}) + \I \Ai(\tilde{z}),\\
  \left(\pdiff[2]{}{\tilde{z}} - \tilde{z}\right)G(\tilde{z}, \tilde{z}') &= \delta(\tilde{z}-\tilde{z}'),\\
  \psi_a(\tilde{z}) &= \frac{2m\Omega\xi^2}{\hbar^2}\int\d{\tilde{z}'}\; G(\tilde{z}, \tilde{z}')\psi_b(\tilde{z}').
\end{align*}
\textit{(The homogeneous contributions vanish if there are no obstructions below the injection site.)}
Note that the output is coherent~\cite{Lee:2015} and has a smooth density dependence $n_a(z) = \abs{\psi_a(z)}^2 \propto 1/\sqrt{-z}$ as remarked in Fig.~1 of~\cite{Schneider:1999} and seen in~\cite{Robins:2001} -- there are no density oscillations of the type indicated in Fig.~3 of~\cite{Edwards:1999}.

For $z<0$, the qualitative form can be deduced from the \gls{WKB} approximation:

\begin{gather*}
  \psi_{\mathrm{WKB}}(z) \propto \frac{1}{\sqrt{p(z)}}e^{S(z)/\I\hbar}, \\
  z(t) = - \frac{gt^2}{2}, \qquad
  p(z) = -mgt = -m\sqrt{-2gz}\\
  S(z) = \int_0^{t}\left(\frac{p^2}{2m} - mgz(t)\right)\d{t}
       = \frac{-mg^2t^3}{3} 
       = -\hbar \frac{2}{3}\sqrt{\frac{-z^3}{\xi^3}},\\
  \psi_{\mathrm{WKB}}(z) \propto \frac{1}{\abs{z}^{1/4}}
  \exp\left(\frac{2\I}{3}\sqrt{\frac{-z^3}{\xi^3}}\right)
\end{gather*}

\subsection*{Interferometry}
To qualitatively understand the interferometry of our setup, we make a few approximations.
We first assume that the pulses are short so that we can effectively treat them as instantaneous.

\paragraph{Ramsey Imaging}
Let the first $\pi/2$ Ramsey pulse happen at time $t_1$ and the second $\pi/2$ Ramsey pulse at time $t_2 = t_1 + t_{\mathrm{wait}}$.
We then ignore the complication that different states take slightly different trajectories so that we can consider the particles as a single two-state system with the two falling states $\ket{\ua}$ and $\ket{\ub}$.

Immediately prior to the pulse at $t_1$, the falling atoms are in state $\ket{\psi_{t_1^{-}}}=\ket{\ua}$ (we use the notation $t^{\pm}_1 = t_1 \pm \epsilon$ for small $\epsilon$).
After the first $\pi/2$ pulse, the state is $\ket{\psi_{t_1^{+}}}\propto \ket{\ua} + \ket{\ub}$ with suitably defined axes.
The states now fall until $t=t_2$ accumulating phases $\theta^{12}_{\ua}$ and $\theta^{12}_{\ub}$ respectively: $\ket{\psi_{t_2^{-}}}\propto \exp(\I\theta^{12}_{\ua})\ket{\ua} + \exp(\I\theta^{12}_{\ub})\ket{\ub}$.
After the second $\pi/2$ pulse, we have
\begin{align}
  \ket{\psi_{t_2^+}} &\propto
  e^{\I\theta^{12}_{\ua}}\bigl(\ket{\ua} + \ket{\ub}\bigr)
  + e^{\I\theta^{12}_{\ub}}\bigl(-\ket{\ua} + \ket{\ub}\bigr)\nonumber\\
  &\propto
  \Bigl(e^{\I\theta^{12}_{\ua}} - e^{\I\theta^{12}_{\ub}}\Bigr)\ket{\ua}
  +
  \Bigl(e^{\I\theta^{12}_{\ua}} + e^{\I\theta^{12}_{\ub}}\Bigr)\ket{\ub}.
\end{align}
The two output channels thus give rise to the following interference patterns:
\begin{subequations}
  \label{eqS:interferometer}
  \begin{align}
    n^{\text{Ramsey}}_{\ua} &\propto 1 - \cos\Bigl(\theta^{12}_{\ua} - \theta^{12}_{\ub}\Bigr),\\
    n^{\text{Ramsey}}_{\ub} &\propto 1 + \cos\Bigl(\theta^{12}_{\ua} - \theta^{12}_{\ub}\Bigr).
  \end{align}
\end{subequations}
Using two-component notation with upper component $\ket{\ua}$ and lower component $\ket{\ub}$, the $\pi/2$ pulses (about the $y$ axis) have matrix form
\begin{gather}
  \mat{U}_{\pi/2}
  =\frac{1}{\sqrt{2}}
  \begin{pmatrix}
    1 & -1\\
    1 & 1
  \end{pmatrix}.
\end{gather}
We can thus summarize the procedure as
\begin{alignat*}{4}
  &\begin{pmatrix}
    1\\
    0
  \end{pmatrix}
  \overset{\pi/2}{\longrightarrow}
  &&\frac{1}{\sqrt{2}}
  \begin{pmatrix}
    1\\
    1
  \end{pmatrix}
  \longrightarrow
  &&\frac{1}{\sqrt{2}}
  \begin{pmatrix}
    e^{\I\theta^{12}_{\ua}}\\
    e^{\I\theta^{12}_{\ub}}
  \end{pmatrix}
  \overset{\frac{\pi}{2}}{\longrightarrow}
  &&\frac{1}{2}
  \begin{pmatrix}
    e^{\I\theta^{12}_{\ua}} - e^{\I\theta^{12}_{\ub}}\\
    e^{\I\theta^{12}_{\ua}} + e^{\I\theta^{12}_{\ub}}
  \end{pmatrix}.\\
  &t=t_{1}^{-} &
  &t=t_{1}^{+} &
  &t=t_{2}^{-} &
  &t=t_{2}^{+}
\end{alignat*}

\paragraph{Spin-Echo Imaging}
We can use a similar notation to consider the spin-echo procedure with an additional $\pi$-pulse at time $t_1 < t_e < t_2$:
\begin{gather}
  \mat{U}_{\pi} =
  \begin{pmatrix}
    0 & -1\\
    1 & 0
  \end{pmatrix}.
\end{gather}
\begin{alignat*}{4}
  &\begin{pmatrix}
     1\\
     0
   \end{pmatrix}
  \overset{\pi/2}{\longrightarrow}
  &&\frac{1}{\sqrt{2}}
     \begin{pmatrix}
       1\\
       1
     \end{pmatrix}
  \longrightarrow
  &&\frac{1}{\sqrt{2}}
     \begin{pmatrix}
       e^{\I\theta^{1e}_{\ua}}\\
       e^{\I\theta^{1e}_{\ub}}
     \end{pmatrix}
  \overset{\pi}{\longrightarrow}
  &&\frac{1}{\sqrt{2}}
     \begin{pmatrix}
       -e^{\I\theta^{1e}_{\ub}}\\
       e^{\I\theta^{1e}_{\ua}}
     \end{pmatrix}\longrightarrow\\
  &t=t_{1}^{-} &
  &t=t_{1}^{+} &
  &t=t_{e}^{-} &
  &t=t_{e}^{+}
\end{alignat*}
\begin{alignat*}{2}
  \longrightarrow
  &\frac{1}{\sqrt{2}}
    \begin{pmatrix}
      -e^{\I(\theta^{1e}_{\ub} + \theta^{e2}_{\ua})}\\
      e^{\I(\theta^{1e}_{\ua} + \theta^{e2}_{\ub})}
    \end{pmatrix}
  \overset{\frac{\pi}{2}}{\longrightarrow}
  &&\frac{1}{2}
     \begin{pmatrix}
       -e^{\I(\theta^{1e}_{\ub} + \theta^{e2}_{\ua})}-e^{\I(\theta^{1e}_{\ua} + \theta^{e2}_{\ub})}\\
       -e^{\I(\theta^{1e}_{\ub} + \theta^{e2}_{\ua})}+e^{\I(\theta^{1e}_{\ua} + \theta^{e2}_{\ub})}
     \end{pmatrix}.\\
  &t=t_{2}^{-} &
  &t=t_{2}^{+}
\end{alignat*}

This gives the following interference patterns:
\begin{subequations}
  \label{eqS:spin-echo}
  \begin{align}
    n^{\text{spin-echo}}_{\ua} &\propto
    1 + \cos\Bigl((\theta^{1e}_{\ub} + \theta^{e2}_{\ua}) - (\theta^{1e}_{\ua} + \theta^{e2}_{\ub})\Bigr),\\
    n^{\text{spin-echo}}_{\ub} &\propto
    1 - \cos\Bigl((\theta^{1e}_{\ub} + \theta^{e2}_{\ua}) - (\theta^{1e}_{\ua} + \theta^{e2}_{\ub})\Bigr).
  \end{align}
\end{subequations}

\paragraph{Impulse Approximations}
To gain further insight, we make the approximation that the pulses and $t_{\mathrm{wait}}$ are short enough that the atoms fall a negligible amount while the potentials are imprinted.
We call this the impulse approximation, because we can neglect the kinetic energy contribution to the action.
Under this approximation, we accumulate the following phases:
\begin{gather}
  \theta^{12}_{a} = \frac{S^{12}}{\hbar}
  \approx \frac{1}{\hbar}\int_{t_1}^{t_2}\!\!\! -V_{a}\bigl(z(t)\bigr)\d{t}
  \approx \frac{t_1-t_2}{\hbar}V_{a}\left(\frac{z_1 + z_2}{2}\right).
\end{gather}
Thus, the interferometer \cref{eqS:interferometer} qualitatively measures the difference $\delta V(z) = V_{\ua}(z) - V_{\ub}(z)$ between the two potentials:
\begin{gather}
  n^{\text{Ramsey}}_{\ua} \propto
  1 - \cos\Bigl(\frac{t_{\mathrm{wait}}}{\hbar}\delta V(z)\Bigr)
\end{gather}
where $z$ is the location of the particles at $t\approx t_1\approx t_2$.
This allows us to define the following dimensionless quantity which roughly characterizes the number of maxima (fringes) expected in the interference pattern, as demonstrated in \cref{fig:powersequence_with_zero}:
\begin{gather}
  N_{\mathrm{fringes}} = \frac{t_{\mathrm{wait}}\;\delta V_{\max}}{2\pi \hbar}
\end{gather}
where $\delta V_{\max} = \max_{z}\delta V(z)$ is the maximum of the differential potential.

Relaxing the impulse approximation, the interference pattern will be smeared, stretched, and will fall as the particles continue to accelerate downwards between the pulse sequences and the imaging time.
In principle, this motion can be backed out to provide direct interferometric tomography of the potentials, but much higher accuracy can be obtained by directly fitting the potential.

If we consider the spin-echo imaging with $t_e = (t_1 + t_2)/2$ in the middle, then \cref{eqS:spin-echo} gives the following interference pattern:
\begin{gather}
    n^{\text{spin-echo}}_{\ua} \propto
    1 + \cos\Biggl(\frac{t_{\mathrm{wait}}\Bigl(\delta V(z_{1e}) - \delta V(z_{e2})\Bigr)}{2\hbar}
      \Biggr)\nonumber\\
    \approx
    1 + \cos\Biggl(\frac{t_{\mathrm{wait}}^2\vect{p}\cdot\vect{\nabla}\delta V(z)}{4m\hbar}\Biggr).
\end{gather}
Here $z_{1e}$ and $z_{e2}$ are approximately the midpoints between the $\pi/2$ and $\pi$ pulses, and are separated by time $t_{\mathrm{wait}}/2$, giving a finite difference between the potentials that effectively differentiates the potential in the direction of propagation $\vect{p}$, again subject to the smearing, stretching, and falling as the impulse approximation is relaxed.
This is clearly visible in the spin-echo imaging \cref{fig:spinechob} where the gradient gives two lobes compared to the Ramsey imaging in \cref{fig:spinechoa} which outlines the potential difference itself.

Relaxing the requirement of weak potentials, the pre-factor in \gls{WKB} approximation will become important, and the interference pattern will start to lose contrast.
Additionally, shadows and transverse focusing will affect the amplitude of the final pattern: the full fitting process properly includes these effects far from the turning point, allowing for a more precise extraction of the potential parameters.

\subsection*{Phase Retrieval}
If the form of the potential is not available for accurate fitting as discussed above, one can provide direct interferometric tomography of the potential using phase retrieval techniques~\cite{Bruning:1974,  Juarez_Salazar:2018, Schwiegerling:2014}. 

The essential idea is express the images $I_{n}(\vect{x})$ in terms of the probability density
\begin{subequations}
  \begin{gather}
    n_{\theta}(\vect{x}) = \bigl\lvert\psi_{1}(\vect{x}) + e^{\I\theta}\psi_{2}(\vect{x})\bigr\rvert^2
  \end{gather}
  where $\psi_{i}(\vect{x})$ are the interfering wavefunctions in the \gls{WKB} approximation,
  \begin{gather}
    \psi_{i}(\vect{x}) = A_{i}(\vect{x})e^{S_{i}(\vect{x})/\I\hbar}
  \end{gather}
  and $\theta$ is a relative phase that can be controlled experimentally.
  Since in our experimental implementation the interferometer is defined by two Ramsey pulses, phase shifting can be effected with very high precision by electronically varying the phase of the second Ramsey pulse.
\end{subequations}

By varying $\theta$, we obtain a set of images $I_n \propto n_{\theta_n}$ with the following form (excluding stochastic noise etc.):
\begin{multline*}
  I_n(\vect{x}) \propto a(\vect{x}) + b(\vect{x})\cos\bigl(\phi(\vect{x})+\theta_n\bigr)\\
  = \underbrace{A_1^2(\vect{x}) + A_2^2(\vect{x})}_{a(\vect{x})} +
  \underbrace{A_1(\vect{x})A_2(\vect{x})}_{b(\vect{x})}
  \cos\Bigl(\underbrace{\frac{S_2(\vect{x}) - S_1(\vect{x})}{\I\hbar}}_{\phi(\vect{x})} + \theta_n\Bigr).
\end{multline*}
Expanding, we have
\begin{gather*}
  I_n(\vect{x}) \propto a(\vect{x}) + b(\vect{x})\cos\phi(\vect{x})\cos\theta_n - b(\vect{x})\sin\phi(\vect{x})\sin\theta_n,
\end{gather*}
thus, the set of images for different $\theta_n$ form an ellipse in a three-dimensional sub-space of the space of images.
Phase retrieval is straightforward, especially if $\theta_n$ can be chosen at will as can be done in our experiment.
Simply choose an equally spaced set of $\theta_n = \left.2\pi n/N\right|_{n=0}^{N-1}$.
Averaging the images gives $b(\vect{x})$, allowing the phase $\phi(\vect{x})$ to be retrieved:
\begin{align}
  b(\vect{x}) &= \frac{1}{N}\sum_{n=0}^{N-1}I_{n}(\vect{x}), &
  \tan \phi(\vect{x}) &= \frac{I_{\theta=0}(\vect{x}) - b(\vect{x})}
                             {I_{\theta=-\pi/2}(\vect{x}) - b(\vect{x})}.
\end{align}
Reduction of stochastic errors can be achieved by using the \gls{SVD} to extract the three principle components (eigenfaces) from the complete set of images.

The phase $\phi(\vect{x})$ directly reconstructs the difference in action between the two trajectories, and the derivatives give access to the differential potential.
To be explicit, for a conservative potential $V(z)$ one has
\begin{align*}
  S(z) &= Et - \int_{z_0}^{z} p(z)\d{z},\\
  S'(z) &= -p(z) = \mp\sqrt{2m(E - V(z))},\\
  \hbar\phi_1'(z) - \hbar\phi_2'(z) &= \sqrt{2m(E - V_1(z))} - \sqrt{2m(E - V_2(z))}.
\end{align*}
Setting the energy $E=0$ at the injection site and expanding $V_i(z) \approx V(z) \pm \delta(z)$ gives:
\begin{gather}
  \delta(z) \approx \frac{\hbar\phi_1'(z) - \hbar\phi_2'(z)}{\sqrt{-2m/V(z)}}.
\end{gather}
Fully inverting $\delta S(x,z)$ for two-dimensional motion is more complicated, but for weak potential differences, everything can be done perturbatively with similar ease.

Our actual experiment is slightly more complicated due to the time-dependence, but the same analysis can be used for a slightly more complicated differential potential of the form
\begin{gather}
  V(z) = \begin{cases}
    V_b(z) & z\in [z_1, z_2],\\
    V_a(z) & \text{otherwise},
  \end{cases}
\end{gather}
where $z_i = z(t_i)$ are the locations of the falling particles when the Ramsey pulses are applied.
The complication is that these locations depend on the unknown differential potential $\delta(z)$, however, for weak potentials, these deviations are small, and the inverse problem can again be solved perturbatively.
Note: this analyses does not assume the impulse approximation.

An analytic expression can be obtained if the differential potential is weak:
\begin{gather}
  S(z_i) \approx S_0(z_i) - \int_{t_1}^{t_2} \delta V\bigl(z(t;z_i)\bigr)\d{t},
\end{gather}
where $z(t)$ is the classical trajectory of the particle that arrives at $z=z(t_i, z_i)$ a the time of imaging.
In this approximation, the contribution of $S_0(z_i)$ will cancel from the interference pattern.
Changing variables to the height $z(t;z_f)$, we have:
\begin{gather}
  \delta S(z_i) = S_2(z_i) - S_1(z_i) \approx  \int_{z_1(t_1;z_i)}^{z_2(t_2;z_i)}
  \frac{\delta V(z)}{\dot{z}(z)}\d{z},
\end{gather}
where we note that, for weak potentials, the velocity $\dot{z}$ depends only on $z$.
\begin{subequations}
Thus, the gradient of the interference pattern is
\begin{gather}
  \delta S'(z_i) \approx
  \left.\frac{\delta V(z)}{\dot{z}(z)}
    \pdiff{z}{z_i}
  \right|_{z_1(t_1;z_i)}^{z_2(t_2;z_i)}.
\end{gather}
For a constant gravitational field $V(z) = mgz$ and $\dot{z} = -g(t-t_0) = - \sqrt{-2gz}$, so
\begin{align}
  z(t; z_i) &= -\frac{g}{2}\Bigl(t-t_0(z_i)\Bigr)^2, &
  \pdiff{z}{z_i} &= \frac{t-t_0(z_i)}{t_i-t_0(z_i)},
\end{align}
where we define $\dot{z}_i=-\sqrt{-2gz_i}$ and
\begin{gather}
  t_0(z_i) = t_i - \sqrt{\frac{-2z_i}{g}}
  = t_i + \frac{\dot{z}_i}{g}.
\end{gather}
Collecting everything:
\begin{gather}
  \delta S'(z_i) \approx
  2\frac{\delta V\bigl(z_1(z_i)\bigr)-\delta V\bigl(z_2(z_i)\bigr)}{\sqrt{-2gz_i}}.
\end{gather}
\end{subequations}
As a check, in the impulse approximation $z_2 \approx z_1 + t_{\text{wait}}\dot{z}$:
\begin{subequations}
\begin{align}
  \delta S'(z_i) &\approx
  \frac{\dot{\bar{z}} t_{\text{wait}}\delta V'(\bar{z})}
       {\dot{z}_i}, &
  \bar{z} &= \frac{z_1 + z_2}{2}.
\end{align}
This is consistent with
\begin{gather}
  \delta S(z_i) = t_{\text{wait}}\delta V\bigl(\bar{z}(z_i)\bigr)
\end{gather}
since
\begin{gather}
  \pdiff{\bar{z}}{z_i} = \frac{\dot{\bar{z}}}{\dot{z}_i}.
\end{gather}
\end{subequations}

\begin{figure}[htbp]
  \centering
  \includegraphics[width=\linewidth]{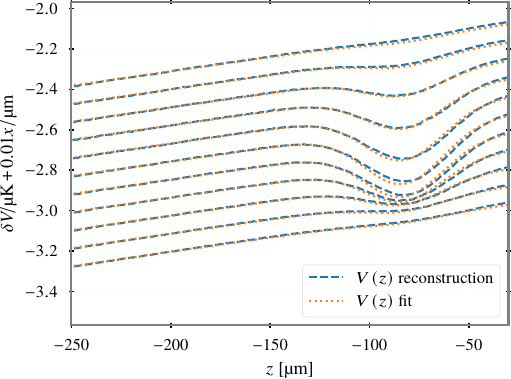}
  \caption{Comparison of the potential differences along the z-axis from phase reconstruction and fits for a number of fixed $x$ positions. Each pair is matched at $z=\SI{-200}{\micro m}$ to compensate for an overall offset in the phase reconstruction algorithm. Inconsistencies where the laser potential is stronger may be due to a small deviation from an ideal Gaussian potential shape, or from a breakdown of the impulse approximation.}
  \label{fig:supp_2}
\end{figure}

As demonstrated in Fig.~\cref{fig:supp_2}, the two methods of reconstructing the differential potential (either from direct experimental phase reconstruction methods or from fitting by matching data using the impulse approximation) result in nearly identical potential shapes and depths over a large area of the atom laser. 
\bibliographystyle{unsrt}
\bibliography{macros,ThesisRef,local}

\exclude{
\iftoggle{mypaper}{
  \iftoggle{arXiv}{
    \bibliography{macros,ThesisRef,local}
  }{}
}{
  \bibliography{macros,ThesisRef,local}
}

\iftoggle{mypaper}{
  \iftoggle{arXiv}{
    \newpage
    \appendix 
  }{ 
    \clearpage
    \appendix
    \setcounter{page}{1}
  }
}{
  \newpage
  \appendix 
}
}

\exclude{
\iftoggle{mypaper}{
  \iftoggle{arXiv}{}{
    \bibliography{macros,ThesisRef,local}}
}{}
}

\end{document}